\documentclass[twocolumn,superscriptaddress, secnumarabic, amssymb, nobibnotes, aps, prd, showpacs]{revtex4-1}
\usepackage{graphicx,subfigure}%
\usepackage{amsmath}

\begin{document}

\title{Simplified scheme for entanglement preparation with Rydberg pumping via dissipation}%

\author{Shi-Lei Su}%
\affiliation{Department of Physics, Harbin Institute of Technology, Harbin 150001, China}
\affiliation{Department of Physics, College
of Science, Yanbian University, Yanji, Jilin 133002, China}

\author{Qi Guo}%
\affiliation{Department of Physics, Harbin Institute of Technology, Harbin 150001, China}
\affiliation{Department of Physics, College
of Science, Yanbian University, Yanji, Jilin 133002, China}

\author{Hong-Fu Wang}%
\affiliation{Department of Physics, College
of Science, Yanbian University, Yanji, Jilin 133002, China}

\author{Shou Zhang}%
\email[ Corresponding author: ]{szhang@ybu.edu.cn}
\affiliation{Department of Physics, Harbin Institute of Technology, Harbin 150001, China}
\affiliation{Department of Physics, College
of Science, Yanbian University, Yanji, Jilin 133002, China}

\begin{abstract}
Inspired by recent work [A. W. Carr and M. Saffman, Phys. Rev. Lett. \textbf{111}, 033607 (2013)],
we propose a simplified scheme to prepare the two-atom maximally entangled states via dissipative Rydberg pumping.
Compared with the former scheme, the simplified one involves less classical laser fields and Rydberg interactions,
and the asymmetric Rydberg interactions are avoided.
Master equation simulations demonstrate that the fidelity and the Clauser-Horne-Shimony-Holt correlation
of the maximally entangled state could reach up to 0.999 and 2.821, respectively, under certain conditions.
Furthermore, we extend the physical thoughts to prepare the three-dimensional entangled state,
the numerical simulations show that, in theory, both the fidelity and the Negativity of the desired entanglement
could be very close to unity under certain conditions.


\end{abstract}
\maketitle

\section{introduction}
The strong and long-range interaction between Rydberg excited atoms~\cite{tf1994,rhj2012,mtk2010}
is suitable for creating magnetic phases with long-range order~\cite{hh2010,hmi2010,emm2011,i2011,sci2011,wmt2012},
and can give rise to the Rydberg blockade that
suppress resonant optical excitation of multiple Rydberg atoms~\cite{djp2000,tmj2006,dsj2004,kmt2004,jrh2011,ylf2012}.
The Rydberg blockade offers many possibilities for realization of the neutral-atom-based
quantum information processing~(QIP) tasks~\cite{mtk2010} and for observing the multi-atom
effects~\cite{jia2010,arg2011,rsk2011,mmj2011,i2012,pmm2012,dka2013}.
Recently, the blockade effect between a single Rydberg-excited atom and a second one located about
10 $\mu$m and 4 $\mu$m away have been observed in experiments~\cite{ett2009,ayt2009}, respectively.
On the other hand, when the strength of the Rydberg interaction is under the intermediate regime, i.e., too weak
to yield the blockade, yet too strong to be ignored, the atoms could be pumped to Rydberg excited states simultaneously.
In this case, the Rydberg-interaction-based controlled-phase gate has been studied theoretically~\cite{djp2000,dk2014}.
Besides, if the Rydberg interaction strength
meets some certain conditions with the detuning of the driving laser fields,
it could also be used to resonantly pump the ground state to the two-excitation state
of two Rydberg atoms. This Rydberg pumping process has theoretically been studied and used for preparation of entanglement~\cite{am2013,xjt2014}.

It has traditionally been considered that the dissipation only has negative effect on performance of QIP tasks.
However, with the development of quantum information technology, researchers had a better insight on the roles of dissipation.
In 1999, Plenio \emph{et al}. and Cabrillo \emph{et al}. first proposed schemes to prepare entanglement via dissipation~\cite{msa1999,cjp1999}.
Immediately afterwards, several schemes were suggested to study the entanglement in dissipative
quantum systems~\cite{sget2002}.
These works have opened new chapters for the dissipative-dynamics-based schemes~\cite{saa2008,hck2011,jmp2011,dal2011,mfa2011}.
Compared with the unitary-dynamics-based schemes for entanglement preparation, the chief features of these dissipative-dynamics-based ones are that
the desired state can be achieved without requirement of setting specific initial state and controlling evolution time accurately.

Schemes that combine dissipative dynamics and Rydberg interactions are interesting since
the advantages of dissipation and that of Rydberg atom are integrated together~\cite{arb2012,am2013}.
In particular, Carr and Saffman proposed a scheme for preparation of two-atom
Bell singlet state with Rydberg state mediated interaction via dissipation~\cite{am2013}.
In contrast with the Rydberg blockade based schemes that accommodate at most one atom shared Rydberg
excitation, the scheme proposed in Ref.~\cite{am2013} utilizes the Rydberg pumping process
as well as the microwave field to drive the undesired states in the ground state subspace to
that in two-excitation subspace which would decay to the desired state in ground state subspace via dissipation with a certain probability.
However, to prevent the pumping process from the desired state to the two-excitation state, asymmetric Rydberg coupling strengths are required.
The scheme in Ref.~\cite{am2013} was extended to prepare three-dimensional entanglement in Ref.~\cite{xjt2014},
in which three Rydberg excited states in each atom are required and thus nine Rydberg interactions would be produced.
Nevertheless, similar to the scheme in Ref.~\cite{am2013}, asymmetric Rydberg coupling strengths are also necessary
since three out of the nine Rydberg interactions are undesirable.

In this paper, getting inspirations from Ref.~\cite{am2013},
we propose a simplified scheme for preparation of
two-atom entanglement by dissipative Rydberg pumping.
Our scheme has the following characteristics:
(i) Only one Rydberg-Rydberg interaction term is produced and thus the asymmetric
Rydberg coupling strengths are avoided.
(ii) Numerical simulations show that the fidelity $\mathcal{F}$ = 0.999 could be achieved.
(iii) The physical thought is exteded to prepare three-dimensional entanglement
through adding one Rydberg-Rydberg interaction without requirement of asymmetry Rydberg interactions.
Compared with the unitary-dynamics-based schemes, the present one has the benefits
such as the desired entanglement could be prepared without requirement of setting
initial state and controlling evolution time accurately.

The organization of the rest paper is as follows: In Sec.~\ref{s2}, we propose the scheme to
prepare two-atom maximal entanglement and assess the performance through calculating
the fidelity and Clauser-Horne-Shimony-Holt correlation numerically. In Sec.~\ref{s3}, we generalize the scheme to
prepare three-dimensional entanglement and assess the performance by calculating the
fidelity and negativity. Discussion and Conclusion are given out in
Secs.~\ref{s4} and \ref{s5}, respectively.

\section{preparation of two-atom Bell state}\label{s2}
\subsection{Basic Model}
\begin{figure}[htp!]
  \includegraphics[scale = 0.5]{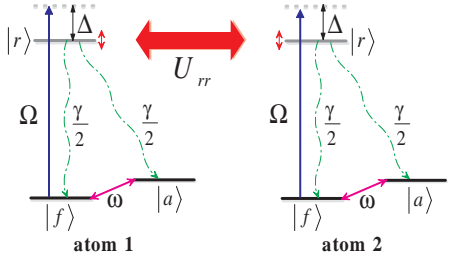}\\
  \caption{(Color online) Energy-level diagram of Rydberg atoms. $|f\rangle$ and $|a\rangle$ denote two ground states, and $|r\rangle$ denotes the Rydberg excited state.
    $U_{rr}$ stands for the strength of the Rydberg interaction.
  $\Delta$ and $\Omega$ denote the detuning and the Rabi frequency of the classical laser field, respectively.
  The resonant transition between two ground states is realized by microwave field or Raman transition with Rabi frequency $\omega$.}\label{f001}
\end{figure}

Our system for dissipative preparation of two-atom Bell state is shown in Fig.~\ref{f001}, which is composed of two atoms. Each of the two atoms
has two ground states $|f\rangle$ and $|a\rangle$ and one Rydberg excited state $|r\rangle$. The transition $|f\rangle\rightarrow|r\rangle$
is driven by the classical optical laser with Rabi frequency $\Omega$ and detuning $-\Delta$. The resonant coupling between ground states
$|f\rangle$ and $|a\rangle$ is realized by microwave field or Raman transition
with Rabi frequency $\omega$. $U_{rr}$ denotes the strength of the Rydberg interaction
which mainly depends on the principal quantum numbers of the Rydberg atom and the distance between the Rydberg atoms. Here, we assume the
excited state can spontaneously decay into two ground states with branching ratio $\gamma/2$.
Thus, in a rotating frame, the Hamiltonian of the system could be written as~(setting $\hbar$ = 1 throughout this paper)
\begin{equation}\label{e001}
\mathcal{\hat{H}} = \mathcal{\hat{H}}_{1}\otimes\mathcal{\hat{I}}_{2} + \mathcal{\hat{I}}_{1}\otimes\mathcal{\hat{H}}_{2} + \mathcal{\hat{V}},
\end{equation}
in which the footnote \emph{j} (\emph{j} = 1, 2) denotes the \emph{j}-th atom, and in the basis \{$|f\rangle$, $|a\rangle$, $|r\rangle$\} the single atom operators
can be expressed as
\begin{eqnarray}\label{e002}
    \mathcal{\hat{H}}_{j} =
    \left(
      \begin{array}{ccc}
        0 & \omega/2 & \Omega/2 \\
        \omega^{*}/2 & 0 & 0 \\
        \Omega^{*}/2 & 0 & -\Delta \\
      \end{array}
    \right),~~
    \mathcal{\hat{I}}_{j} =
    \left(
      \begin{array}{ccc}
        1 & 0 & 0 \\
        0 & 1 & 0 \\
        0 & 0 & 1 \\
      \end{array}
    \right).
\end{eqnarray}
The Rydberg interaction energy can be written as
\begin{equation}\label{e003}
\mathcal{\hat{V}} = U_{rr}~|r\rangle_{11}\langle r|\otimes|r\rangle_{22}\langle r|.
\end{equation}
The Lindblad operator describing the decay induced by the atomic spontaneous emission can
be expressed as $\mathcal{\hat{L}}_{1}=\sqrt{\gamma/2}|f\rangle_{11}\langle r|$,
$\mathcal{\hat{L}}_{2}=\sqrt{\gamma/2}|a\rangle_{11}\langle r|$,
$\mathcal{\hat{L}}_{3}=\sqrt{\gamma/2}|f\rangle_{22}\langle r|$,
$\mathcal{\hat{L}}_{4}=\sqrt{\gamma/2}|a\rangle_{22}\langle r|$.
Thus, the master equation that describing the evolution of the whole system can be written as
\begin{equation}\label{e004}
    \mathcal{\dot{\hat{\rho}}} = i[\hat{\rho}, \mathcal{\hat{H}}] + \frac{1}{2}\sum_{k=1}^{4}\big[2\mathcal{\hat{L}}_{k}\hat{\rho}\mathcal{\hat{L}}_{k}^{\dag}
    -(\mathcal{\hat{L}}_{k}^{\dag}\mathcal{\hat{L}}_{k}\hat{\rho}+\hat{\rho}\mathcal{\hat{L}}_{k}^{\dag}\mathcal{\hat{L}}_{k})\big].
\end{equation}

\subsection{Unitary Dynamics and Dissipative Dynamics}\label{s2.2}

In order to identify the roles of unitary dynamics and dissipative dynamics concisely, we here
use the triplet-singlet basis of the ground states: $\{|ff\rangle,
~|S\rangle = (|fa\rangle - |af\rangle)/\sqrt{2},~ |T\rangle = (|fa\rangle + |af\rangle)/\sqrt{2},~ |aa\rangle\}$,
in which $|mn\rangle$ is the abbreviation of $|m\rangle_{1}|n\rangle_{2}$ and we shall conform throughout this paper
to this notation for simplify. $|S\rangle$ is the desired state we want to prepare.
Driven by the classical laser field and the Rydberg interaction,
transformations of the states can be divided into the following three types.
(i) $|S\rangle$ and $|T\rangle$ would be excited to
$(|ra\rangle - |ar\rangle)/\sqrt{2}$ and $(|ra\rangle + |ar\rangle)/\sqrt{2}$, respectively, with detuning $-\Delta$.
(ii) $|ff\rangle$ would be excited to $|fr\rangle$ or $|rf\rangle$ with detuning $-\Delta$ and be further excited to
$|rr\rangle$ with detuning $-2\Delta + U_{rr}$.
(iii) $|aa\rangle$ keeps invariant. Thus, it is easy to see from type~(ii) that the selection of $U_{rr} = 2\Delta$
would lead to a resonant coupling between $|ff\rangle$ and $|rr\rangle$~(this process is the so-called Rydberg pumping).
Meanwhile, if the parameters satisfy $\Delta\gg\Omega$ (large detuning), the states
in single excitation subspace can be safely discarded. That is, the transformations
described in type~(i) can be ignored. According to Eq.~(\ref{e002}), the Hamiltonian
of microwave could be written as~(assuming that $\omega$ is real):
\begin{equation}\label{e005}
    \mathcal{\hat{H}_{\omega}} = \frac{\omega}{2}\sum_{j=1,2}(|f\rangle_{j}\langle a| + {\rm H.c.}).
\end{equation}
It is easy to verify that $|S\rangle$ is the dark state of Hamiltonian in Eq.~(\ref{e005}). Nevertheless, the states
$|ff\rangle$, $|T\rangle$ and $|aa\rangle$ would be translated to each other under the control of $\mathcal{\hat{H}_{\omega}}$.
Besides, using the triplet-singlet basis, the Lindblad operators can be reexpressed as
\begin{eqnarray}\label{e006}
    \mathcal{\hat{L}}_{1} &=& \sqrt{\frac{\gamma}{2}}\big[|fr\rangle\langle rr| + (|T\rangle + |S\rangle)\langle ra| + |ff\rangle\langle rf|\big],\cr\cr
    \mathcal{\hat{L}}_{2} &=& \sqrt{\frac{\gamma}{2}}\big[|ar\rangle\langle rr| + (|T\rangle - |S\rangle)\langle rf| + |aa\rangle\langle ra|\big],\cr\cr
    \mathcal{\hat{L}}_{3} &=& \sqrt{\frac{\gamma}{2}}\big[|rf\rangle\langle rr| + (|T\rangle - |S\rangle)\langle ar| + |ff\rangle\langle fr|\big],\cr\cr
    \mathcal{\hat{L}}_{4} &=& \sqrt{\frac{\gamma}{2}}\big[|ra\rangle\langle rr| + (|T\rangle + |S\rangle)\langle fr| + |aa\rangle\langle ar|\big].
\end{eqnarray}
Thus, it is easy to get that the two-excitation state $|rr\rangle$ would be decayed to the states in single excitation subspace
which would be further decayed to the states in the subspace composed by $\{|ff\rangle,|S\rangle, |T\rangle, |aa\rangle\}$.

The unitary dynamics and the dissipative dynamics described above enable the desired state $|S\rangle$ being the unique steady state of the system.
In other words, if the initial state is $|S\rangle$, it keeps invariant.
If the initial state is one of the other three states or their superposition state in the ground state subspace,
it would be excited to $|rr\rangle$ directly or indirectly via the Rydberg pumping process and transition process induced by microwave field,
and then decay to the state in the ground-state subspace via dissipation.
The final state would be re-excited and re-decayed again until the desired state is prepared.

\subsection{Performance of the scheme}

\subsubsection{Fidelity}
The most common way to assess the quality of the steady state is fidelity.
Usually, the fidelity of steady state is defined as $\mathcal{\hat{F}} = \langle \psi|\hat{\rho}_{t\rightarrow \infty}|\psi\rangle$,
in which $|\psi\rangle$ is the desired state and $\hat{\rho}_{t\rightarrow \infty}$ denotes the practical steady state density matrix.
\begin{figure}[htp!]
  \includegraphics[scale = 0.5]{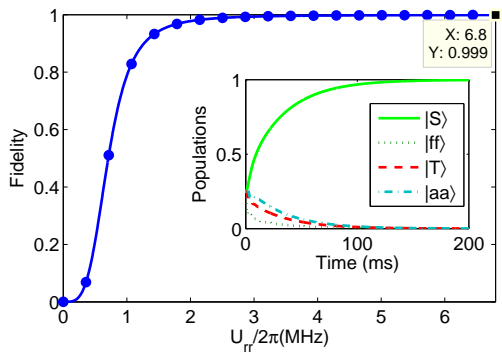}\\
  \caption{(Color online) Fidelity of steady-state $|S\rangle$ with respect to the Rydberg interaction strength $U_{rr}$.
  The rest parameters are chosen as $\Omega/2\pi$=0.036~MHz, $\gamma=1.673$~KHz, $\omega \simeq 0.004\Omega$, and $\Delta=U_{rr}/2$.
  The inset shows the evolution of populations versus time when initial state is
  the mixed state $(|ff\rangle\langle ff| + |S\rangle\langle S| + |T\rangle\langle T| + |aa\rangle\langle aa|)/4$
  for $\Delta/2\pi$ = 3.435~MHz.}\label{f002}
\end{figure}
In Fig.~\ref{f002}, we plot the fidelity of state $|S\rangle$ for initial state $|ff\rangle$.
The inset of Fig.~\ref{f002} shows the evolution of populations versus time, from which one could see
that the scheme is initial-state-independent and does not require controlling evolution time accurately.

\subsubsection{Clauser-Horne-Shimony-Holt correlation}
\begin{figure}[htp!]
  \includegraphics[scale = 0.5]{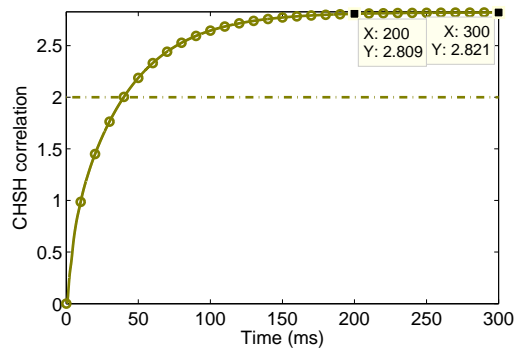}\\
  \caption{(Color online) The CHSH correlation of steady-state $|S\rangle$ with respect to evolution time.
  The rest parameters are chosen as
  $\Omega/2\pi$=0.036~MHz, $\gamma=1.673$~KHz, $\omega \simeq 0.004\Omega$, $\Delta/2\pi$ = 3.435~MHz, and $U_{rr}=2\Delta$.
  The initial state is the mixed state
  $(|ff\rangle\langle ff| + |S\rangle\langle S| + |T\rangle\langle T| + |aa\rangle\langle aa|)/4$.}\label{f003}
\end{figure}

The premises of locality and realism imply some constraints
on the statistics of two spatially separated particles, which are
known as Bell inequalities~\cite{bell1965}.
On that basis,  Clauser, Horne, Shimony, and Holt derived the Clauser-Horne-Shimony-Holt~(CHSH)~correlation~\cite{jma1969}.
The system state violates Bell inequality when the CHSH correlation rises above 2. And the quantum mechanisms predicts CHSH correlation $\mathcal{B}(t)$
equals $2\sqrt{2}$ for the maximal violation limit. In Fig.~\ref{f003}, we plot the CHSH correlation
\begin{equation}\label{e007}
    \mathcal{B}(t) = {\rm Tr}[(\mathcal{O}_{\rm CHSH})\rho(t)]
\end{equation}
with respect to evolution time.
In Eq.~(\ref{e007}), $\mathcal{O}_{\rm CHSH}$ is defined as
\begin{eqnarray}\label{e008}
\mathcal{O}_{\rm CHSH} &=& \sigma_{y,1}\otimes\frac{-\sigma_{y,2}-\sigma_{x,2}}{\sqrt{2}}
                         +\sigma_{x,1}\otimes\frac{-\sigma_{y,2}-\sigma_{x,2}}{\sqrt{2}}\cr\cr&&
                         +\sigma_{x,1}\otimes\frac{\sigma_{y,2}-\sigma_{x,2}}{\sqrt{2}}
                         -\sigma_{y,1}\otimes\frac{\sigma_{y,2}-\sigma_{x,2}}{\sqrt{2}}.
\end{eqnarray}
From Fig.~\ref{f003}, one can see that the stable CHSH correlation 2.821 close to the maximal
violation limit $2\sqrt{2}\approx2.828$, which resulting in a constant violation of Bell inequality.

\subsection{Preparation of state $|T\rangle$}

The former scheme can be easily generalized to prepare $|T\rangle$ through adding a $\pi$ relative phase of
the microwave field for the two Rydberg atoms. On that basis the corresponding Hamiltonian of microwave field
in the singlet-triplet basis can be written as
\begin{equation}\label{e009}
    \mathcal{\hat{H}_{\omega}} = \frac{\omega}{2}\sum_{j=1,2}(e^{(j-1)i\pi}|f\rangle_{j}\langle a| + {\rm H.c.}),
\end{equation}
which has no effect on $|T\rangle$ while translates $|ff\rangle, |S\rangle$ and $|aa\rangle$ with each other.
Combining with the repeated Rydberg pumping and dissipative processes, the state $|T\rangle$ would be prepared
in a way similar to the former scheme. With parameters the same to the former scheme, numerical simulations
show that the optimal values of fidelity and the CHSH correlation can also reach 0.999 and 2.821, respectively.

\section{preparation of three-dimensional entangled state}\label{s3}

\subsection{Basis Model}

\begin{figure}[htp!]
  \includegraphics[scale = 0.5]{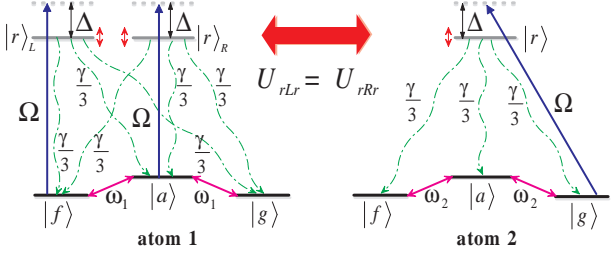}\\
  \caption{(Color online) Energy-level diagram of Rydberg atoms. $|f\rangle$, $|a\rangle$ and $|g\rangle$ denote three ground states, and $|r\rangle$ denotes the Rydberg excited state.
    $U_{rL(R)r}$ stands for the strength of the Rydberg interaction.
  $\Delta$ and $\Omega$ denote the detuning and the Rabi frequency of the classical laser field, respectively.
  The resonant transition $|f\rangle\leftrightarrow|a\rangle$ and $|a\rangle\leftrightarrow|g\rangle$ is realized by microwave or Raman transition with Rabi frequency $\omega_{1}$ and $\omega_{2}$, respectively. $\gamma/3$ is the spontaneous emission rate. For simplicity, we have assumed the spontaneous emission rates from Rydberg state to different ground states equal to each other.}\label{f004}
\end{figure}

The scheme to prepare three-dimensional entangled state is illustrated in Fig.~\ref{f004}.
We consider two cesium atoms, both of which have three ground states
$|f\rangle\equiv|F=3,M=-1\rangle$, $|a\rangle\equiv|F=4,M=0\rangle$ and $|g\rangle\equiv|F=3,M=1\rangle$ of the 6$S_{1/2}$ mainfold.
Atom 1 has two Rydberg states $|r\rangle_{L}\equiv|F=3,M=-1\rangle$ and $|r\rangle_{R}\equiv|F=4,M=0\rangle$ of 125$P_{1/2}$
while atom 2 has one Rydberg state $|r\rangle\equiv|F=4,M=1\rangle$ of 125$P_{1/2}$.
According to the notations of Fig.~\ref{f004}, the Hamiltonian in a rotating frame can be written as
\begin{equation}\label{e010}
\mathcal{\hat{H}} = \mathcal{\hat{H}}_{1}\otimes\mathcal{\hat{I}}_{2} + \mathcal{\hat{I}}_{1}\otimes\mathcal{\hat{H}}_{2} + \mathcal{\hat{V}}.
\end{equation}
In the basis $\{|f\rangle, |a\rangle, |g\rangle, |r\rangle_{L}, |r\rangle_{R}\}$, $\mathcal{\hat{H}}_{1}$ can be written as
\begin{eqnarray}\label{e011}
    \mathcal{\hat{H}}_{1} =
    \left(
      \begin{array}{ccccc}
        0 & \omega_{1}/2 & 0 & \Omega/2& 0 \\
        \omega_{1}^{*}/2 & 0 & \omega_{1}/2 & 0 & \Omega/2\\
        0 & \omega_{1}^{*}/2 & 0 & 0 & 0\\
          \Omega^{*}/2 & 0 & 0 &  -\Delta & 0\\
            0 & \Omega^{*}/2 & 0 & 0 &  -\Delta\\
      \end{array}
    \right).
\end{eqnarray}
And $\mathcal{\hat{I}}_{1}$ is $5\times5$ identical matrix.
In the basis $\{|f\rangle, |a\rangle, |g\rangle, |r\rangle\}$, $\mathcal{\hat{H}}_{2}$ can be written as
\begin{eqnarray}\label{e012}
    \mathcal{\hat{H}}_{2} =
    \left(
      \begin{array}{ccccc}
        0 & \omega_{2}/2 & 0 & 0\\
        \omega_{2}^{*}/2 & 0 & \omega_{2}/2 & 0\\
        0 & \omega_{2}^{*}/2 & 0 & \Omega/2 \\
         0& 0& \Omega^{*}/2 &  -\Delta \\
      \end{array}
    \right).
\end{eqnarray}
$\mathcal{\hat{I}}_{2}$ is $4\times4$ identical matrix. And the Rydberg interaction terms can be expressed as
\begin{equation}\label{e013}
\mathcal{\hat{V}} = U_{r_{L}r}~|r\rangle_{LL}\langle r|\otimes|r\rangle_{22}\langle r| + U_{r_{R}r}~|r\rangle_{RR}\langle r|\otimes|r\rangle_{22}\langle r|.
\end{equation}
In the following, we assume $U_{r_{L}r} = U_{r_{R}r} = U_{rr}$. That is, the asymmetry Rydberg interactions are not required.
The nine Lindblad operators induced by the dissipation is
$\mathcal{\hat{L}}_{r_{L},f(a,g)}=\sqrt{\gamma/3}|f(a,g)\rangle_{11}\langle r_{L}|$,
$\mathcal{\hat{L}}_{r_{R},f(a,g)}=\sqrt{\gamma/3}|f(a,g)\rangle_{11}\langle r_{R}|$,
$\mathcal{\hat{L}}_{r,f(a,g)}=\sqrt{\gamma/3}|f(a,g)\rangle_{22}\langle r|$.
The master equation thus can be written as
\begin{equation}\label{e014}
\mathcal{\dot{\hat{\rho}}} = i[\hat{\rho}, \mathcal{\hat{H}}] + \frac{1}{2}\sum_{k}\big[2\mathcal{\hat{L}}_{k}\hat{\rho}\mathcal{\hat{L}}_{k}^{\dag}
    -(\mathcal{\hat{L}}_{k}^{\dag}\mathcal{\hat{L}}_{k}\hat{\rho}+\hat{\rho}\mathcal{\hat{L}}_{k}^{\dag}\mathcal{\hat{L}}_{k})\big].
\end{equation}

\subsection{Unitary dynamics and dissipative dynamics}\label{s3.2}

In this subsection, for clearness we redefine the basis in the ground state subspace as
\{$|fa\rangle,|fg\rangle,|af\rangle,|ag\rangle,|gf\rangle,|ga\rangle,|\phi\rangle=(|ff\rangle+|aa\rangle+|gg\rangle)/\sqrt{3},
|\varphi\rangle=(|ff\rangle-2|aa\rangle+|gg\rangle)/\sqrt{6},|\psi\rangle=(|ff\rangle-|gg\rangle)/\sqrt{2}$\}, in which $|\phi\rangle$
is the desired three dimensional entanglement. Similar to the principles in Sec.~\ref{s2.2},
if the parameters satisfy $2\Delta = U_{rr}\gg\Omega$, Rydberg pumping processes,
$|f\rangle|g\rangle\rightarrow|r\rangle_{L}|r\rangle$ and $|a\rangle|g\rangle\rightarrow|r\rangle_{R}|r\rangle$,
become the main transition process
induced by classical field and Rydberg interactions since the transition processes
$|f\rangle|f\rangle\rightarrow|r\rangle_{L}|f\rangle$,
$|f\rangle|a\rangle\rightarrow|r\rangle_{L}|a\rangle$,
$|a\rangle|f\rangle\rightarrow|r\rangle_{R}|f\rangle$,
$|a\rangle|a\rangle\rightarrow|r\rangle_{R}|a\rangle$ and
$|g\rangle|g\rangle\rightarrow|g\rangle|r\rangle$
are detuned by $\Delta$. Besides, $|g\rangle|f\rangle$ and $|g\rangle|a\rangle$ are free from the influence of classical field.

Based on Eqs.~(\ref{e011}) and~(\ref{e012}), the Hamiltonian of microwave field can be written as
\begin{eqnarray}\label{e015}
\mathcal{\hat{H}_{\omega}} = \frac{\omega}{2}(|f\rangle_{1}\langle a|
+ |a\rangle_{1}\langle g|-|f\rangle_{2}\langle a| - |a\rangle_{2}\langle g|) + {\rm H.c.}.~~~~
\end{eqnarray}
One could easily verify that $|\phi\rangle$ is the dark state of Hamiltonian shown in Eq.~(\ref{e015}).
Meanwhile, the other eight states in the redefined basis would be translated to each other under the control of Eq.~(\ref{e015}),
in which we have chosen $\omega_{1} = -\omega_{2} =\omega$ and $\omega$ has been considered as reals.

The Lindblad operators can also be rewritten as:
\begin{eqnarray}\label{e016}
  \mathcal{\hat{L}}_{r_{L(R)},f} &=& \sqrt{\frac{\gamma}{3}}\Big[|fr\rangle\langle r_{L(R)}r|
  + |fa\rangle\langle r_{L(R)}a| + |fg\rangle\langle r_{L(R)}g|
  \cr\cr&&+(\frac{1}{\sqrt{3}}|\phi\rangle + \frac{1}{\sqrt{6}}|\varphi\rangle + \frac{1}{\sqrt{2}}|\psi\rangle)\langle r_{L(R)}f|\Big],
 \cr\cr \mathcal{\hat{L}}_{r_{L(R)},a} &=& \sqrt{\frac{\gamma}{3}}\Big[|ar\rangle\langle r_{L(R)}r|
  + |af\rangle\langle r_{L(R)}f| + |ag\rangle\langle r_{L(R)}g|
  \cr\cr&&(\frac{1}{\sqrt{3}}|\phi\rangle - \frac{\sqrt{6}}{3}|\varphi\rangle)\langle r_{L(R)}a|\Big],
  \cr\cr\mathcal{\hat{L}}_{r_{L(R)},g} &=& \sqrt{\frac{\gamma}{3}}\Big[|gr\rangle\langle r_{L(R)}r|
  + |gf\rangle\langle r_{L(R)}f| + |ga\rangle\langle r_{L(R)}a|
  \cr\cr&&+(\frac{1}{\sqrt{3}}|\phi\rangle + \frac{1}{\sqrt{6}}|\varphi\rangle - \frac{1}{\sqrt{2}}|\psi\rangle)\langle r_{L(R)}g|\Big],
  \cr\cr\mathcal{\hat{L}}_{r,f} &=& \sqrt{\frac{\gamma}{3}}\Big[|r_{L}f\rangle\langle r_{L}r|+|r_{R}f\rangle\langle r_{R}r|
  + |af\rangle\langle ar|
  \cr\cr&&+ |gf\rangle\langle gr|+(\frac{1}{\sqrt{3}}|\phi\rangle + \frac{1}{\sqrt{6}}|\varphi\rangle
  + \frac{1}{\sqrt{2}}|\psi\rangle)\langle fr|\Big],
  \cr\cr \mathcal{\hat{L}}_{r,a} &=& \sqrt{\frac{\gamma}{3}}\Big[|r_{L}a\rangle\langle r_{L}r| + |r_{R}a\rangle\langle r_{R}r|
  + |fa\rangle\langle fr|
  \cr\cr&&+|ga\rangle\langle gr|+(\frac{1}{\sqrt{3}}|\phi\rangle - \frac{\sqrt{6}}{3}|\varphi\rangle)\langle ar|\Big],
  \cr\cr\mathcal{\hat{L}}_{r,g} &=& \sqrt{\frac{\gamma}{3}}\Big[|r_{L}g\rangle\langle r_{L}r|+|r_{R}g\rangle\langle r_{R}r|
  + |ag\rangle\langle ar|
  \cr\cr&&+ |fg\rangle\langle fr|+(\frac{1}{\sqrt{3}}|\phi\rangle + \frac{1}{\sqrt{6}}|\varphi\rangle
  - \frac{1}{\sqrt{2}}|\psi\rangle)\langle gr|\Big].~~~~
\end{eqnarray}
It is easy to see that two-excitation states $|r_{L(R)}r\rangle$ would be decayed to
single excitation states, which would be further decayed to the states in the ground state subspace.

The Rydberg pumping process as well as the states transition process induced by the microwave fields resonantly
drive the undesired state in the ground state subspace to the two-excitation Rydberg state, which would
decay to the desired state via dissipation with a specific probability. However, when $|r_{L(R)}r\rangle$
is transformed to the undesired state, it would be re-drive and re-decay again until $|\phi\rangle$ is prepared.

\subsection{Performance of the scheme}

\subsubsection{Fidelity}
\begin{figure}
  \includegraphics[scale=0.5]{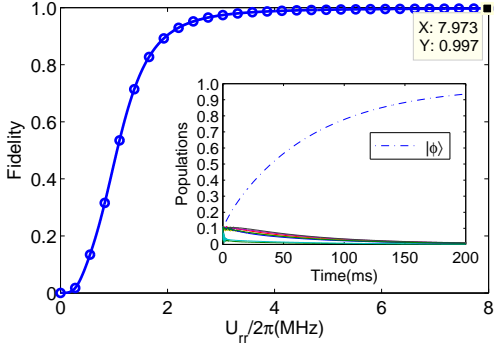}\\
  \caption{(Color online) Fidelity versus the Rydberg interaction strength $U_{rr}$.
  The rest parameters are chosen as $\Omega/2\pi$ = 0.055~MHz, $\omega$ = 0.0075$\Omega$, $\gamma$ = 1~KHz and $\Delta$ = $U_{rr}/2$.
  The inset shows the populations of all of the states in the basis defined in Sec.~\ref{s3.2} when $\Delta/2\pi$ = 2~MHz with respect to time.
  The initial state is mixed state and chosen as ($\sum_{p}|p\rangle\langle p|$)/9, in which $|p\rangle$ is any one of the
  states in the basis defined in Sec.~\ref{s3.2}.
  }\label{f005}
\end{figure}

Figure~\ref{f005} shows the steady state fidelity of the desired three dimensional entangled state $|\phi\rangle$ with respect to Rydberg interaction strength for the parameters $\Omega/2\pi$ = 0.055~MHz,
$\omega$ = 0.0075$\Omega$, $\gamma$ = 1~KHz and $\Delta = U_{rr}/2$. It is clear to see that the fidelity increases with increasing $U_{rr}$.
The inset of Fig.~\ref{f005} indicates that the population of desired state would be higher than 91\% with the evolution time 200 ms.

\subsubsection{Negativity}

Negativity is a measure deriving by Vidal and Werner from the Peres' criterion for separability~\cite{GRF2002},
which is defined as
\begin{equation}\label{e017}
    \mathcal{N}(\hat{\rho}_{AB}) = \frac{||\hat{\rho}_{A,B}^{T_{A}}||-1}{2},
\end{equation}
where $\hat{\rho}_{A,B}^{T_{A}}$ means the partial transpose of the
bipartite mixed state $\rho_{A,B}$ with respect to the subsystem \emph{A}, and
\begin{equation}\label{e018}
    ||\hat{\rho}_{A,B}^{T_{A}}|| = {\rm Tr}|\hat{\rho}_{A,B}^{T_{A}}| = {\rm Tr}\sqrt{\hat{\rho}_{A,B}^{T_{A}\dag}\hat{\rho}_{A,B}^{T_{A}}}
\end{equation}
is the trace norm of the operator $\rho_{A,B}^{T_{A}}$.
\begin{figure}
  \includegraphics[scale=0.5]{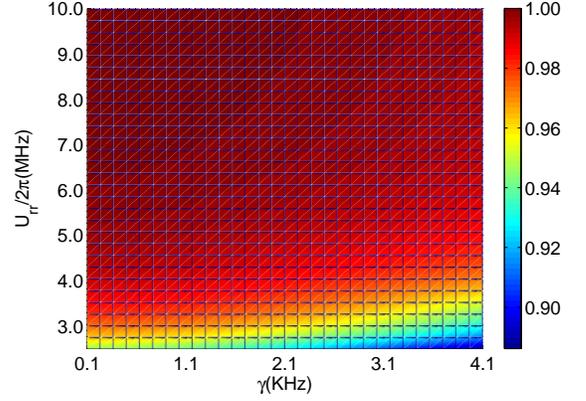}\\
  \caption{(Color online) Negativity versus the Rydberg interaction strength $U_{rr}$ and the spontaneous emission rate $\gamma$.
  The rest parameters are chosen as $\Omega/2\pi$ = 0.055~MHz, $\Delta$ = $U_{rr}/2$ and $\omega$ = 0.0075$\Omega$}\label{f006}
\end{figure}
An alternative and equivalent definition of negativity is the absolute sum of the negative eigenvalues of $\hat{\rho}_{A,B}^{T_{A}}$~\cite{PAM1998}
\begin{equation}\label{e019}
     \mathcal{N}(\hat{\rho}_{AB}) = \sum_{j}\frac{|\lambda_{j}| - \lambda_{j}}{2},
\end{equation}
where $\lambda_{j}$ are all of the eigenvalues of $\hat{\rho}_{A,B}^{T_{A}}$. Ideally, the value of the negativity of a pure high dimensional
entangled state equals 1.
Through solving the master equation~(\ref{e014}) numerically, we plot the negativity of steady state with respect to Rydberg interaction strength and
the atomic spontaneous emission rate in Fig.~\ref{f006}, which shows that the negativity is higher than 90\% over most of the range of parameters.
In particular, the value 0.9995 is found when $\Delta/2\pi = 4.8705$ MHz and $\gamma$ = 1.033~KHz.

\subsection{Generalization of the scheme}

The scheme can be used to prepare $|\phi^{'}\rangle=(|ff\rangle-|aa\rangle+|gg\rangle)/\sqrt{3}$ via modulating the
Rabi frequency of microwave field as $\omega_{1} = \omega_{2} =\omega$. Under other fixed conditions, the variant of
Rabi frequency of microwave field makes $|\phi'\rangle$ being its dark state and thus be prepared via Rydberg pumping
process and dissipative process. Numerical simulations show that the fidelity and negativity could also be higher than 99\%.

\section{Discussion}\label{s4}

Unlike traditional Rydberg interaction based schemes, the current one utilizes the interaction to pump
some specific states rather than to constitute the blockade mechanism. To achieve this,
a certain condition between the detuning of the driving field and the strength of Rydberg coupling should be
fulfilled. The combination of transition processes induced by microwave field and pumping process induced by Rydberg interaction drives
the undesired state to the two-excitation Rydberg state resonantly, which would decay to
the desired state via dissipation with a specific probability. Once the two-excitation Rydberg state
decays to the undesired state, it would be re-drive and re-decay again.
The whole process is more like the automatic repeat-until-success mechanism since it has no need for measurement and
intervention.

\begin{figure}
  \includegraphics[scale=0.7]{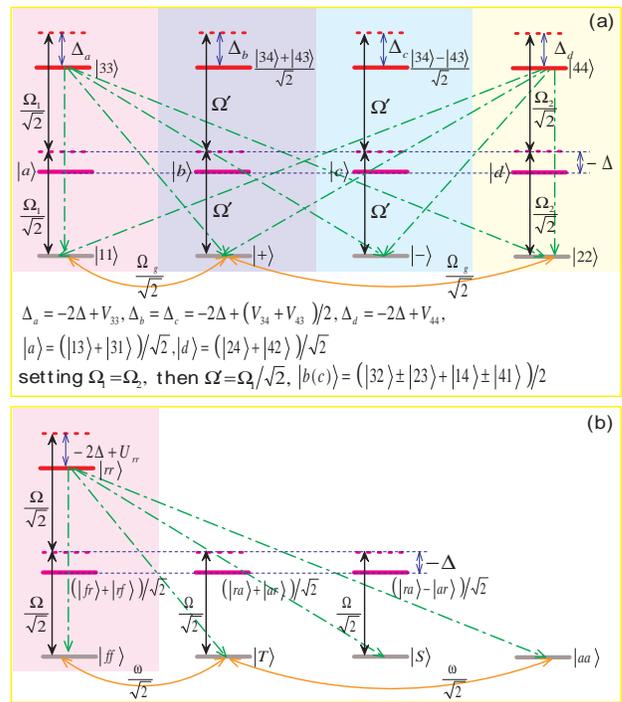}\\
  \caption{(a) Diagram of the effective coupling and dissipative processes in the two-atom basis
  extracted from the scheme proposed by  Carr and Saffman~\cite{am2013}. For clearness, the symbols are used the same as Ref.~\cite{am2013}.
  The dash-and-dot lines
  denote the decay induced by atomic spontaneous emission. In order to pump the undesired states
  to the two-excitation subspace and protect the desired state from pumping processes,
  the four Rydberg interaction strengths should be adjusted to
  satisfy $V_{33}=V_{44}=2\Delta, V_{34}=V_{43}=\Delta$, and the detuning should satisfy $\Delta\gg{\Omega_{1},\Omega_{2}}$.
  In this case, $\Delta_{a}=\Delta_{d}=0$, therefore,
  the pumping processes $|11\rangle\rightarrow|33\rangle$ and $|22\rangle\rightarrow|44\rangle$ are preserved,
  while the pumping processes $|+\rangle\rightarrow(|34\rangle+|43\rangle)/\sqrt{2}$ and $|-\rangle\rightarrow(|34\rangle-|43\rangle)/\sqrt{2}$
  are discarded. With the help of the microwave-field-induced transitions, the desired state $|-\rangle$ would be prepared.
  (b) Diagram of the effective coupling and dissipative processes in the two-atom basis
  extracted from one of our schemes. To achieve the desired state, the Rydberg interaction strength
  and the detuning should fulfill $U_{rr}=2\Delta$ and $\Delta\gg\Omega$, respectively.
  By comparing (a) and (b), one could see that our scheme involves
 less laser beams and less Rydberg interaction strength that should
 be adjusted.}\label{f007}
\end{figure}

To avoid the undesirable Rydberg pumping processes which would drive the desired state to the two-excitation Rydberg state,
the weak asymmetric Rydberg interactions are introduced in Ref.~\cite{am2013}.
Although the asymmetry Rydberg interaction could be achieved when the Rydberg excited states corresponding to
different Zeeman sublevels~\cite{tm2008}, it would no doubt add experimental complexity.
Without any performance deterioration, our simplified scheme
involves only one Rydberg interaction rather than four and thus the asymmetry Rydberg interactions are avoided.
Furthermore, the simplified scheme involves less laser beams. In order to indicate the characteristics of the present and the former scheme,
we plot the effective coupling processes of these two schemes in Fig.~\ref{f007}(a) and Fig.~\ref{f007}(b), respectively.

In the process of generalizing the scheme to prepare three-dimensional entanglement,
 the energy-level diagram we first considered is not the same as Fig.~\ref{f004}.
 Only one Rydberg state is considered in atom 1 and thus only one Rydberg pumping process is used to prepare the entanglement.
However, in this case, the numerical simulation indicates that the optical fidelity of the desired state is not higher than 43.2\%.
This situation could be interpreted as follows: For the scheme proposed in Sec.~\ref{s2},
three undesired ground states should be pumped to the two-excitation subspace via
Rydberg pumping process directly or indirectly~(with the help of transition process induced by microwave field).
Nevertheless, for the scheme proposed in Sec.~\ref{s3},
there are eight undesired states in the ground state subspace which should be pumped to the two-excitation subspace.
It is more difficult to pump eight rather than three states via only one Rydberg pumping process.
In order to improve this situation, we re-design the scheme with two Rydberg states in atom 1.
The Rydberg pumping process $|fg\rangle\rightarrow|r_{L}r\rangle$ and $|ag\rangle\rightarrow|r_{R}r\rangle$
as well as the microwave-fields-induced transition process drive all of the eight undesired states to the two-Rydberg states effectively.
But even so, the asymmetry Rydberg interactions are still not required since the Rydberg interaction strengths in Fig.~\ref{f004} satisfy $U_{r_{L}r}=U_{r_{R}r}$.
Numerical simulations of the master equations indicate that
the three-dimensional entanglement can be prepared with fidelity and negativity being higher than 0.997 and 0.999, respectively.
Compared with the scheme proposed in Ref.~\cite{xjt2014}, the scheme in Sec.~\ref{s3}
(i) involves less classical laser fields.
(ii) involves two Rydberg pumping processes rather than nine and the symmetry Rydberg interactions are avoided.

\begin{figure}
  \includegraphics[scale = 0.6]{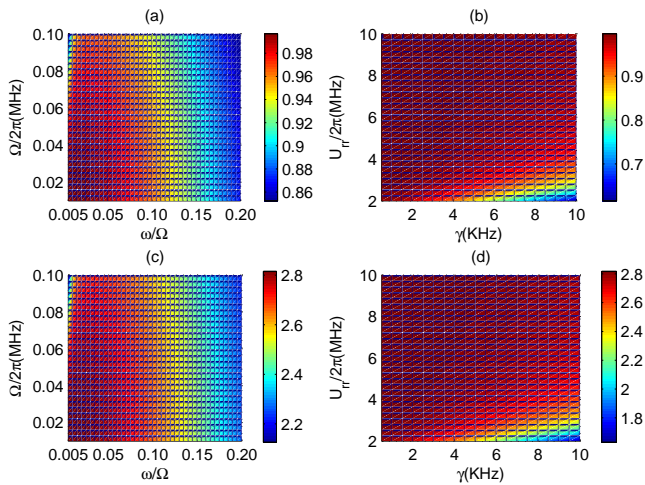}\\
  \caption{(Color online) Robustness on parameters variation of the scheme proposed in Sec.~\ref{s2}.
  (a)((c)): Fidelity (CHSH correlation) of state $|S\rangle$ with respect to $\Omega$ and $\omega$
  with $\gamma$ = 1~KHz, $U_{rr}/2\pi$ = 4~MHz and $\Delta = U_{rr}/2$.
  (b)((d)): Fidelity (CHSH correlation) of state $|S\rangle$ with respect to $U_{rr}$ and $\gamma$
  with $\Omega/2\pi$ = 0.036~MHz, $\omega \approx 0.004\Omega$ and $\Delta = U_{rr}/2$.}\label{f008}
\end{figure}

\begin{figure}
  \includegraphics[scale = 0.6]{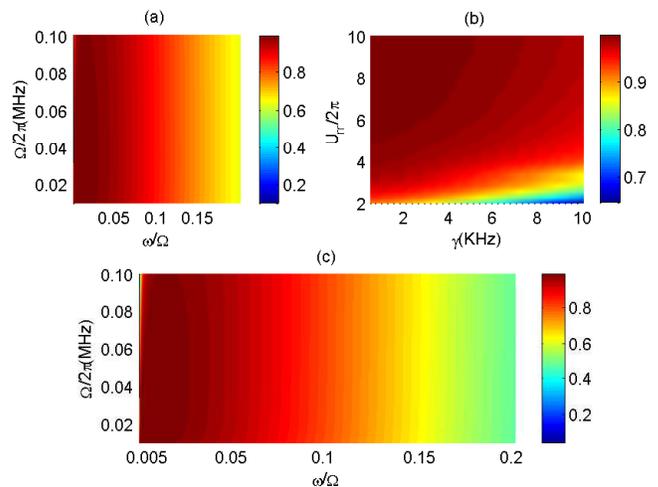}\\
  \caption{(Color online) Robustness on parameters variation of the scheme proposed in Sec.~\ref{s3}.
  (a)((c)) Fidelity (Negativity) of state $|\phi\rangle$ with respect to $\Omega$ and $\omega$
  with $\gamma$ = 1~KHz, $U_{rr}/2\pi$ = 6~MHz and $\Delta = U_{rr}/2$. (b) Fidelity of state
  $|\phi\rangle$ with respect to $U_{rr}$ and $\gamma$
  with $\Omega/2\pi$ = 0.055~MHz, $\omega = 0.0075\Omega$ and $\Delta = U_{rr}/2$.}\label{f009}
\end{figure}

We now briefly discuss the experimental feasibility of the parameters, such as $\Omega$, $\omega$ and $\gamma$.
In theory, the one-photon Rabi frequency $\Omega$ is associated with the field amplitude and polarization,
which could be adjusted under current technology~\cite{mtk2010}.
Experimentally, $\Omega/2\pi$=0.49~MHz~\cite{tet2008}, 0.67~MHz~\cite{lex2010} and (0.1, 3.2)~MHz~\cite{jda2010} for the Rydberg
atoms have been achieved and used for QIP tasks.
If the coupling among ground states is realized by the microwave field, $\omega$ could be adjusted in a way similar to $\Omega$.
Otherwise, if it is realized by the effective Raman process via the medium state~\cite{dj2007},
$\omega$ could be adjusted through modifying the coupling between the ground states and the medium state.
The lifetime of the excited state of Rydberg atoms has been investigated in Ref.~\cite{iid2009},
from which one could get that $\gamma=1.43,1.67,2$~KHz can be obtained for
Rb or Cs atom at the appropriate temperature.
We now analyze the robustness of the scheme on parameters' variation. In Fig.~\ref{f008},
we plot fidelity and CHSH correlation of steady-state proposed in Sec.~\ref{s2}
versus different parameters, which show that the fidelity and CHSH correlation
would be higher than 0.9 and 2.2 within most of the range of parameters. Analogously,
in Fig.~\ref{f009}, we plot fidelity and negativity of steady-state proposed
in Sec.~\ref{s3} versus different parameters, from which one could see that three-dimensional
state could also be prepared with high fidelity and negativity over most of the range of parameters.
It is necessary to mention that, in this paper, the numerical solutions of the master equations is performed with
the Quantum Optics Toolbox with Matlab~\cite{sm1999}.

It should be noted that although the theoretical analysis show that the scheme
could put up a good performance, some realistic situations may influence the
scheme in practical experiments. Since the current scheme simplifies and generalizes the one proposed in Ref.~\cite{am2013}, the
discussions of feasibility of the former scheme could also fit our scheme well.
For instance, as pointed out in the former scheme, spontaneous emission in real atoms may populate
other hyperfine ground states outside the encoded ground states, and this situation could be dealt with by using recycling lasers.
Besides, if the Rydberg interactions are considered spin-dependent, the undesired entanglement between
spin and center of mass degrees of freedom may also decrease the performance of the scheme.
However, this can be suppressed if operators confine the atoms in the Lamb-Dicke regime and
use magic ground-Rydberg trapping potentials~\cite{am2013,sfm2011}.

\section{Conclusion}\label{s5}

In conclusion, we have put forward an efficient scheme to prepare two-atom maximal entanglement via dissipative Rydberg
pumping process. Then, the scheme was generalized to prepare three-dimensional entanglement.
The steady states could be achieved with less resources and lower experimental requirements.
Moreover, compared with the unitary dynamics based schemes,
our scheme has no requirement of initial states and controlling evolution time accurately.

\begin{center}
{\bf{ACKNOWLEDGMENT}}
\end{center}
This work was supported by the National Natural Science Foundation
of China under Grant Nos. 11264042, 11465020 and 61465013.

\end{document}